# Inverse Problems for Matrix Exponential in System Identification: System Aliasing

Zuogong Yue, Johan Thunberg* and Jorge Gonçalves

*Abstract*— This note addresses identification of the $A$-matrix in continuous time linear dynamical systems on state-space form. If this matrix is partially known or known to have a sparse structure, such knowledge can be used to simplify the identification. We begin by introducing some general conditions for solvability of the inverse problems for matrix exponential. Next, we introduce "system aliasing" as an issue in the identification of slow sampled systems. Such aliasing give rise to non-unique matrix logarithms. As we show, by imposing additional conditions on and prior knowledge about the $A$-matrix, the issue of system aliasing can, at least partially, be overcome. Under conditions on the sparsity and the norm of the $A$-matrix, it is identifiable up to a finite equivalence class.

## I. INTRODUCTION

Time-series models in engineering, economics and biology can often be represented by state-space models. In the example of gene regulatory networks (evolving close to equilibria), the structure of the right-hand side defines pathways, from which conclusions can be drawn about possible diseases. For continuous-time linear dynamics, the state-space form has an $A$-matrix, which reveals the direct connections between the states. In many dynamical systems, as the aforementioned ones, the structure of this matrix is either partially known or known to be sparse. This paper investigates what such criteria can be used to identify the $A$-matrix.

Estimating continuous-time systems from discrete-time measured data is an important part of the field of systems identification, see e.g. [1]–[3]. However, with low sampling rates, the identification of continuous-time systems becomes particularly challenging, manifested in the lack of comprehensive studies. In the presence of "system aliasing", the discrete-time signals do not contain certain information about the continuous-time signals. As a result, even though the discrete-time system can be identified, the selection of the continuous-time model may be ambiguous. This note sheds light on how this issue of system aliasing complicates the identification.

Before addressing the issues related to system aliasing, we first recall some results on matrix logarithms and exponentials – the questions of existence and uniqueness are addressed. The results are given as algebraic conditions for obtaining a unique $A$-matrix within a set. In contrast to these results, the note proceeds by first providing the minimal sampling frequency such that system aliasing is avoided.

This work was supported by Fonds National de la Recherche Luxembourg (9247977 & 8864515).
All authors are with Luxembourg Centre for Systems Biomedicine (LCSB), University of Luxembourg, 6, avenue du Swing, L-4367 Belvaux, Luxembourg.
*For correspondence, johan.thunberg@uni.lu

Then we consider the issue of system aliasing. When know that the $A$-matrix is sparse, we observe that allowing for "aliased" representations might lead to $A$-matrices that are more sparse among the aliased solutions, thus exposing more structures in the state generation. This note gives a mathematical definition of "system aliases", and study how to select the sparest one among the aliases of the underlying systems. Refer to [4] for more preliminaries and all proofs.

Let
$$A = \bar{A} + E \text{ and } D \in \mathbb{R}^{p \times n^2},$$
where $\bar{A} \in \mathbb{R}^{n \times n}$ and $E \in \mathscr{S} \subseteq \mathbb{R}^{n \times n}$. This note addresses properties that must hold for $(A, D, \mathscr{S})$ or $(\bar{A}, E, D, \mathscr{S})$ in order to guarantee that $E$ can be determined from $D\text{vec}(\exp(A))$. In the following cited definitions and theorems, we adopt the notations in [5].

### A. Principal logarithm

**Theorem 1** (principal logarithm [5, Thm. 1.31]). *Let $P \in \mathbb{C}^{n \times n}$ have no eigenvalues on $\mathbb{R}^-$. There is a unique logarithm $X$ of $P$ all of whose eigenvalues lie in the strip $\{z : -\pi < \text{im}(z) < \pi\}$. We refer to $X$ as the principal logarithm of $P$ and write $X = \text{Log}(P)$. If $P$ is real then its principal logarithm is real.*

Let $\mathcal{G}(h) = \{z \in \mathbb{C} : -\pi/h < \text{im}(z) < \pi/h, h \in \mathbb{R}\}$. We denote the set of real matrices in $\mathbb{R}^{n \times n}$ whose eigenvalues lie in the strip $\mathcal{G}(1)$ by $\mathscr{A}(n)$. By restricting the set for which $\bar{A}$ and $E$ belong to $\mathscr{A}(n)$, it follows that
$$E = \text{Log}(\exp(A)) - \bar{A},$$
is one-to-one. Throughout the text, the notations $\exp(\cdot)$ and $e^{(\cdot)}$ are used interchangeably. We use $\log(\cdot)$ for general *primary* matrix logarithms and $\text{Log}(\cdot)$ for principal logarithms.

**Theorem 2** (Gantmacher [5, Thm. 1.27]). *Let $P \in \mathbb{C}^{n \times n}$ be nonsingular with the Jordan canonical form*

$$Z^{-1}PZ = J = \text{diag}(J_1, J_2, ..., J_p) \tag{1a}$$

$$J_k = J_k(\lambda_k) = \begin{bmatrix} \lambda_k & 1 & & \\ & \lambda_k & \ddots & \\ & & \ddots & 1 \\ & & & \lambda_k \end{bmatrix} \in \mathbb{C}^{m_k \times m_k}. \tag{1b}$$

*Then all solutions to $e^A = P$ are given by*

$$A = ZU \text{diag}(L_1^{j_1}, L_2^{j_2}, ..., L_p^{j_p})U^{-1}Z^{-1}, \tag{2}$$

*where*
$$L_k^{j_k} = \log(J_k(\lambda_k)) + 2j_k\pi i I_{m_k}; \quad (3)$$

$\log(J_k(\lambda_k))$ *denotes*

$$f(J_k) \coloneqq \begin{bmatrix} f(\lambda_k) & f'(\lambda_k) & \cdots & \frac{f^{(m_k-1)}(\lambda_k)}{(m_k-1)!} \\ & f(\lambda_k) & \ddots & \vdots \\ & & \ddots & f'(\lambda_k) \\ & & & f(\lambda_k) \end{bmatrix}$$

*with $f$ the principal branch of the logarithm, defined by $\text{Im}(\log(z)) \in (-\pi, \pi]$; $j_k$ is an arbitrary integer; and $U$ is an arbitrary nonsingular matrix that commutes with $J$.*

**Theorem 3** (classification of logarithms [5, Thm. 1.28]). *Let the nonsingular matrix $P \in \mathbb{C}^{n \times n}$ have the Jordan canonical form (1) with $p$ Jordan blocks, and let $s \leq p$ be the number of distinct eigenvalues of A. Then $e^A = P$ has a countable infinity of solutions that are primary functions of $P$, given by*

$$A_j = Z \, \text{diag}(L_1^{j_1}, L_2^{(j_2)}, ..., L_p^{(j_p)}) Z^{-1}, \quad (4)$$

*where $L_k^{j_k}$ is defined in (3), corresponding to all possible choices of the integers $j_1, ..., j_p$, subject to the constraint that $j_i = j_k$ whenever $\lambda_i = \lambda_k$.*

### B. Fréchet Derivatives

**Definition 1** ( [5]). The *Fréchet derivative* of the matrix function $f : \mathbb{C}^{n \times n} \to \mathbb{C}^{n \times n}$ at a point $X \in \mathbb{C}^{n \times n}$ is a linear map

$$\begin{array}{ccc} \mathbb{C}^{n \times n} & \xrightarrow{L} & \mathbb{C}^{n \times n} \\ E & \longmapsto & L(X, E) \end{array}$$

such that for all $E \in \mathbb{C}^{n \times n}$

$$f(X+E) - f(X) - L(X, E) = o(\|E\|).$$

The Fréchet derivative exists for matrix functions $\exp$ and $\text{Log}$ (principal logarithm) and it is unique. It holds that [5, p. 238]

$$L_{\exp}(X, E) = \int_0^1 e^{X(1-s)} E e^{Xs} ds,$$

$$L_{\text{Log}}(X, E) = \int_0^1 (t(X-I)+I)^{-1} E (t(X-I)+I)^{-1} dt.$$

### C. Fréchet derivatives for the vector representation

The vector representation of the Fréchet derivatives $L_{\exp}(X, E)$ and $L_{\text{Log}}(X, E)$ have the structure of being given as a matrix multiplied by $\text{vec}(E)$.

$$\text{vec}(L_{\exp}(X, E)) = K(X, 0) \text{vec}(E), \quad (5)$$

$$\text{vec}(L_{\text{Log}}(X, E)) = K(X, 0)^{-1} \text{vec}(E), \quad (6)$$

$$K(X, E) = \big(I \otimes \exp(X)\big) \psi\big((X+E)^T \oplus (-X)\big). \quad (7)$$

Here $K(X, E)$ can be seen as an extension of the object K(X) defined in [5, Thm. 10.13]. The vector representation of $e^A = e^{\bar{A}+E}$ can be written as

$$\text{vec}(e^{\bar{A}+E}) = \text{vec}(e^{\bar{A}}) + K(\bar{A}, E) E. \quad (8)$$

Moreover, let us define the maps $f_{\bar{A}} : \mathbb{R}^{n \times n} \to \mathbb{R}^{n^2}$ and $g_{\bar{A}} : \mathbb{R}^{n \times n} \to \mathbb{R}^{n^2}$ by

$$f_{\bar{A}}(E) = K(\bar{A}, E) \text{vec}(E) \quad \text{and} \quad g_{\bar{A}}(E) = K(\bar{A}, 0) \text{vec}(E).$$

## II. PROBLEM FORMULATIONS

### A. General notation

**Definition 2.**

$$\mathcal{E}(A, D, h, \mathscr{S}) = \Big\{ A^* \in \mathbb{R}^{n \times n} : \ D \in \mathbb{R}^{n^2 \times n^2}, h \in \mathbb{R},$$

$$A^* = \arg\min_{\tilde{A} \in \mathscr{S}} \|D\text{vec}(\exp(hA)) - D\text{vec}(\exp(h\tilde{A}))\|_2 \Big\},$$

where $\mathscr{S} \subseteq \mathbb{R}^{n \times n}$ contains $A$.

This general notation will be used to define an important concept of *system aliasing* in continuous-time linear system identification.

Consider the special case of the set $\mathscr{S}$ whose elements $A^*$ are all in $\mathscr{A}(n)$. Moreover, to easily apply Fréchet derivatives, with $A$ expressed as $A = \bar{A} + E$ at $\bar{A}$, we have the following definitions.

**Definition 3.**

$$\mathcal{E}_L(\bar{A}, E, D, \mathscr{S}) = \Big\{ E^* : \bar{A} + E^* \in \mathscr{A}(n), \ D \in \mathbb{R}^{n^2 \times n^2},$$

$$E^* = \arg\min_{\tilde{E} \in \mathscr{S}} \|D\text{vec}(\exp(\bar{A}+E)) - D\text{vec}(\exp(\bar{A}+\tilde{E}))\|_2 \Big\},$$

where $\mathscr{S} \subseteq \mathbb{R}^{n \times n}$ contains $E$.

If $E$ is "sufficiently" small in norm, there is an approximated but easier problem that we could investigate, where we can use Fréchet derivatives to approximate $\exp(\bar{A}+E)$ at $\bar{A}$ in the direction of $E$.

**Definition 4.**

$$\mathcal{E}_S(\bar{A}, E, D, \mathscr{S}) = \Big\{ E^* : \bar{A} + E^* \in \mathscr{A}(n), \ D \in \mathbb{R}^{n^2 \times n^2},$$

$$E^* = \arg\min_{\tilde{E} \in \mathscr{S}} \|DK(\bar{A}, 0)\big(\text{vec}(E) - \text{vec}(\tilde{E})\big)\|_2 \Big\},$$

where $\mathscr{S} \subseteq \mathbb{R}^{n \times n}$ is a linear subspace containing $E$.

Here we use the first order approximation of the exponential matrix. Then we assume that $\mathcal{S}$ is a linear subspace and we want to find out in what directions in this space the first order approximation is good. To be more precise, the question is for what $(\bar{A}, E, D, \mathcal{S})$ it holds that

$$\frac{\left( \sup_{E' \in \mathcal{E}_S(\bar{A}, tE, D, \mathcal{S})} \|E' - tE\| \right)}{|t|} \to 0 \quad \text{as } t \to 0. \quad (9)$$

### B. Continuous-time linear system identification

Consider the linear dynamical system

$$\begin{cases} \dot{x}(t) = Ax(t), \\ y(t) = Cx(t), \end{cases} \quad (10)$$

where $A = \bar{A} + E$; $C \in \mathbb{R}^{p \times n}$ has full rank, $x(t) \in \mathbb{R}^n$ and $y(t) \in \mathbb{R}^p$.

With the general notation given in Section II-A, we can give a definition on *system aliasing* only using the $A$ matrix and the sampling period $h$, which is not necessary to depend on specific identification methods.

**Definition 5** (System aliasing). Given $A \in \mathscr{S}$ and $h \in \mathbb{R}^+$, if there exists $\hat{A} \neq A \in \mathscr{E}(A, I, h, \mathscr{S})$ and $\hat{A}$ is called *system alias* of $A$ with respect to $\mathscr{S}$. By default, choose $\mathscr{S}(A) := \{\tilde{A} \in \mathbb{R}^{n \times n} : \max\{\text{im}(\text{eig}(\tilde{A}))\} \leq \max\{\text{im}(\text{eig}(A))\}\}$.

We are particularly interested in $\mathscr{E}(A, I, h, \mathscr{S}) = \{A\}$, i.e. there is no problem of *system aliasing*. Note that the concept of *system aliasing* does not depend on specific data. It only depends on system dynamics (e.g. the $A$-matrix in (10)) and sampling frequencies. If the $D$ matrix is specifically constructed by data instead of $I$, $\mathscr{E}(A, D, h, \mathscr{S}) = \{A\}$, where $A$ denotes the ground truth, tells that the underlying system is identifiable from the given data (see Section III-B). Obviously if we have system aliasing for the system with a specific sampling frequency, without extra prior information on $A$ (see Section IV), the system is not identifiable.

## III. NO SYSTEM ALIASING

### A. The minimal sampling frequency

Provided with the definition of *system aliasing*, a question comes first: for what $(A, h)$, it holds that $\mathscr{E}(A, I, h, \mathscr{S}(A)) = \{A\}$.

To make principal matrix logarithm $\text{Log}(\cdot)$ well-defined, assume that $\exp(hA)$ has no negative real eigenvalues. By Theorem 1 and 2, it always holds that $\text{Log}(\exp(hA))/h \in \mathscr{E}(A, I, h, \mathscr{S}(A))$. To avoid *system aliasing*, we have to force $\text{Log}(\exp(hA))/h = A$ to be satisfied. It is equivalent to $\text{eig}(hA) \in \mathcal{G}(1)$.

Given no other information on the system, consider the identification problem of $A$ using full-state measurement. The only way to find the unique estimation is to decrease the sampling period $h$ until the ground truth falls into the strip of $\mathcal{G}(h)$, and then use the principal logarithm, as illustrated in Figure 1. Otherwise, we would be bothered by *system aliases* of $A$ and unable to make a decision, unless we know extra prior information on $A$. For full-state measurement, identifiability is guaranteed by selecting appropriate $h$ such that there is no *system aliases*. For the general case of identification using output measurement, the issue is studied in Section III-B.

**Theorem 4** (Nyquist-Shannon-like sampling theorem). *To uniquely obtain $A$ from $A_d$ by taking the principal matrix logarithm, where $A_d$ is identified from sampled data, the sampling frequency $\omega$ (rad/s) must satisfy*

$$\omega \geq 2 \max \{|\,\text{im}\,(\lambda_i(A))\,|,\ i = 1, \ldots, n\}.$$

*Equivalently, the sampling period $h$ should satisfy*

$$h \leq \min \{\pi/|\,\text{im}\,(\lambda_i(A))\,|,\ i = 1, \ldots, n\}.$$

*Proof.* The proof immediately follows from the above explanation on principal matrix logarithms. $\square$

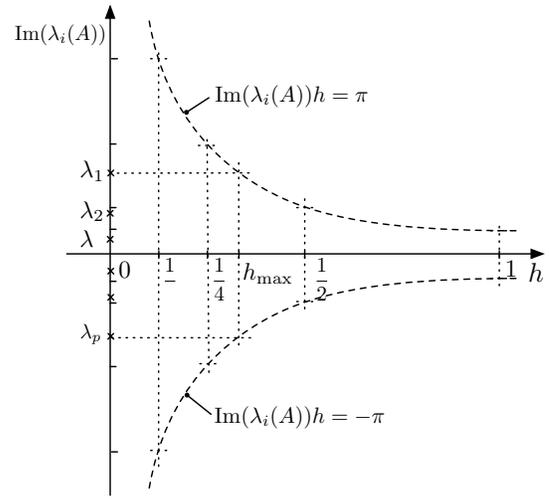

Fig. 1: The imaginary parts of all eigenvalues of $A$ must lie into $(-\pi/h, \pi/h)$. $\lambda_i(\cdot)$ denotes the $i$-th eigenvalue of $A$ in Theorem 2. The symbols "×" denote the locations of $\text{im}(\lambda_i(A))$. $h_{\max}$ is the maximal sampling period that allows taking principal logarithms to estimate $A$, without facing troubles from system aliasing.

### B. Partial information

Suppose all $A$'s in $\mathscr{A}(n)$, which implies there is no system aliasing, i.e. the case in Definition 3. Now consider the identifiability problem of (10) from data with low sampling frequency $2\pi/h$. To be precise, it is to find out for what $(\bar{A}, E, D, \mathscr{S})$, it holds that $\mathscr{E}_L(\bar{A}, E, D, \mathscr{S}) = \{E\}$.

**Lemma 5.** *For $\bar{A}$, $E$, $D$, $\mathscr{S} \ni E$, if there is linear subspace $\mathscr{L}$ such that $f_{\bar{A}}(\mathscr{S}) \subseteq \mathscr{L}$, then $\mathscr{E}_L(\bar{A}, E, D, \mathscr{S}) = \{E\}$ if*

$$\mathscr{L} \cap \ker(D) = \{0\}.$$

**Lemma 6.** *For $\bar{A}$, $E$, $D$ and $\mathscr{S} \ni E$, (9) holds if and only if*

$$g_{\bar{A}}^{-1}(\ker(D)) \cap \mathscr{S} = \{0\}.$$

**Proposition 7.** *If*

$$\mathscr{S} \subseteq \{\tilde{E} : \text{im}(\tilde{E}) \cup \text{im}(\bar{A}\tilde{E}) \cup \text{im}(\bar{A}^2\tilde{E}) \cup \ldots \cup \text{im}(\bar{A}^{n-1}\tilde{E}) \subseteq \text{im}(C^T)\}.$$

*and*

$$D = (X^T \otimes C),$$

*where $X \in \mathbb{R}^{n \times n}$ is a non-singular matrix, then $\mathscr{E}_L(\bar{A}, E, D, \mathscr{S}) = \{E\}$ and*

$$E = \log \left( \begin{bmatrix} C \\ Z \end{bmatrix}^{-1} \begin{bmatrix} Ce^A X \\ Ze^{\bar{A}} X \end{bmatrix} X^{-1} \right) - \bar{A},$$

*where $Z$ is any matrix in $\mathbb{R}^{(n-p) \times n}$ such that $\text{im}(Z^T) = \ker(C)$.*

*Proof.*

$$\text{im}(\int_0^1 e^{\bar{A}(1-s)} E' U(s) ds) \subseteq \text{span}\{E', \bar{A}E', \bar{A}^2 E', \ldots, \bar{A}^{n-1} E'\}.$$

Now, if

$$\mathscr{S} \subseteq \{\tilde{E} : \operatorname{im}(\tilde{E}) \cup \operatorname{im}(\bar{A}\tilde{E}) \cup \operatorname{im}(\bar{A}^2\tilde{E}) \cup \ldots$$
$$\cup \operatorname{im}(\bar{A}^{n-1}\tilde{E}) \subseteq \operatorname{im}(C^T)\},$$

for any $E' \in \mathscr{S}$, it holds that $K(\bar{A}, E')\operatorname{vec}(E') \in \operatorname{im}(I \otimes C^T)$. But $\operatorname{im}(I \otimes C^T) = \operatorname{im}(D^T)$. Now, according to Lemma 5, $\mathscr{E}_L(\bar{A}, E, D, \mathscr{S}) = \{E\}$.

Furthermore, since $\operatorname{im}(\int_0^1 e^{\bar{A}(1-s)} E e^{(\bar{A}+E)s} ds) \subseteq C^T$, $Ze^A = Ze^{\bar{A}}$. It follows that

$$\begin{bmatrix} C \\ Z \end{bmatrix} e^A X = \begin{bmatrix} Ce^A X \\ Ze^{\bar{A}} X \end{bmatrix}$$

□

**Corollary 8.** *If*

$$\mathscr{S} \subseteq \{\tilde{E} : \operatorname{im}(\tilde{E}) \cup \operatorname{im}(\bar{A}\tilde{E}) \cup \operatorname{im}(\bar{A}^2\tilde{E}) \cup \ldots$$
$$\cup \operatorname{im}(\bar{A}^{n-1}\tilde{E}) \subseteq \operatorname{im}(X)\}.$$

*and*

$$D = (C \otimes X^T),$$

*where $p = n$, $C$ is a non-singular matrix and $X \in \mathbb{R}^{n \times k}$ where $k \leq n$, then $\mathscr{E}_L(\bar{A}^T, E^T, D, \mathscr{S}) = \{E\}$ and*

$$A = \log \left( \begin{bmatrix} X^T \\ Z \end{bmatrix}^{-1} \begin{bmatrix} X^T e^{A^T} C^T \\ Ze^{\bar{A}^T} C^T \end{bmatrix} C^{-T} \right)^T,$$

*where $Z$ is any matrix in $\mathbb{R}^{(n-k) \times n}$ such that $\operatorname{im}(Z^T) = \ker(C)$.*

*Proof.* Identify $\bar{A}$ with $\bar{A}^T$, $E$ with $E^T$, $C$ with $X^T$, and $X^T$ with $C$ in Proposition 7. □

*Remark* 1. If the columns of $X$ are linearly independent and each column is an initial point for the system (10), Corollary 8 can be used to provide an upper bound on the number of measurements needed in order to guarantee that $\mathscr{E}_L(\bar{A}, E, D, \mathscr{S}) = \{E\}$.

**Proposition 9.** *Suppose the system (10) is initialized at $k$ different initial points $x_{0,1}, x_{0,2}, \ldots, x_{0,k} \in \mathbb{R}^n$, $C \in \mathbb{R}^{p \times n}$ is a full rank matrix and $\mathscr{S}$ is an $l$-dimensional linear subspace. At the time $t = 1$, for initial point $i$, $y_i(1) = Ce^A x_{0,i}$, where $i \in \{1, 2, \ldots, k\}$.*

*For almost all linearly independent vectors $x_{0,1}, x_{0,2}, \ldots x_{0,k}$ in $\mathbb{R}^n$ and almost all matrices $\bar{A}$ in $\mathbb{R}^{n \times n}$, (9) holds where*

$$D = \begin{bmatrix} x_{0,1} & x_{0,2} & \ldots & x_{0,k} \end{bmatrix}^T \otimes C,$$

*if and only if*

$$l \leq kp.$$

*Proof.* For linearly independent $x_{0,1}, x_{0,2}, \ldots x_{0,k}$,

$$D = \begin{bmatrix} x_{0,1} & x_{0,2} & \ldots & x_{0,k} \end{bmatrix}^T \otimes C$$

has rank equal to $kp$. Let $\{v_1, v_2, \ldots, v_l\}$ be a basis for $\mathscr{S}$. Each $v_i$ can be written as $v_i = v_i^{\ker} + v_i^{\operatorname{im}}$ where $K(\bar{A}, 0)v_i^{\ker} \in \ker(D)$ and $K(\bar{A}, 0)v_i^{\operatorname{im}} \in \operatorname{im}(D^T)$. For almost all linearly independent $x_{0,1}, x_{0,2}, \ldots x_{0,k}$ and $\bar{A}$ it holds that the $v_i^{\operatorname{im}}$ are linearly independent if and only if $l \leq kp$.

$$g_{\bar{A}}^{-1}(\ker(D)) \cap \mathscr{S} = \{0\}$$

if and only if the $v_i$ are linearly independent. Thus, according to Proposition 6, (9) holds. □

*Remark* 2. In Proposition 9, instead of having $k$ different initial points one can have one intial point and sample $y(t)$ at the times $t = 1$, $t = 2$ etc. for almost all $\bar{A}$, $E$ and $x_0$ such that $[x_0, e^A x_0, e^{2A} x_0, \ldots, e^{(k-1)A}]$ has full rank (9) holds when

$$D = (\begin{bmatrix} x_0^T & (e^A x_0)^T & \ldots & (e^{(k-1)A} x_0)^T \end{bmatrix}) \otimes C,$$

$l \leq kp$ and $\mathscr{S}$ is an $l$-dimensional linear subset.

## IV. SYSTEM ALIASING AND BOUNDED CONSTRAINTS

In the previous section we hinted that the conditions on no *system aliasing* follows as a consequence of bounded eigenvalues. In this section we follow this path and explicitly formulate an optimization problem to deal with identification in the presence of *system aliases*.

Consider the case of full-state measurement, i.e. $C = I$ in (10), and the sampling period $h$ is NOT chosen small enough such that $\mathscr{E}(A, I, h, \mathscr{S}(A)) = \{A\}$. Then finding out $A$ among its aliases need extra information, for instance, properties of $A$ known *a priori*. Here assume that the ground truth $A$ is the sparest solution in $\mathscr{E}(A, I, h, \mathscr{S}(\kappa))$ and $\kappa \in \mathbb{R}$ as an upper bound can be roughly estimated, where $\mathscr{S}(\kappa)$ will be defined after giving Definition 6. It implies that $A$ is chosen by solving the following optimization problem

$$\underset{\hat{A} \in \mathscr{E}(A, I, h, \mathscr{S}(\kappa))}{\operatorname{minimize}} \|\hat{A}\|_0. \quad (11)$$

We need to calculate $\mathscr{E}(A, I, h, \mathscr{S}(\kappa))$ from data. Given the measurement $X_1 = [x(h), x(2h), \ldots, x(Nh)]$, $X_2 = [x(0), x(h), \ldots, x((N-1)h)]$, let $\hat{A}_d$ be an estimation of the $A$-matrix in the corresponding discrete-time state space representation. In the deterministic case[1] as (10), $\hat{A}_d = X_1 X_2^T (X_2 X_2^T)^{-1}$ and $\hat{A}_d = \exp(hA)$. Hence,

$$\mathscr{E}(A, I, h, \mathscr{S}(\kappa)) = \left\{ \tilde{A} \in \mathscr{S}(\kappa) : \exp(h\tilde{A}) = \hat{A}_d \right\}, \quad (12)$$

and define

$$\mathcal{S} := \left\{ \tilde{A} \in \mathbb{R}^{n \times n} : \exp(h\tilde{A}) = \hat{A}_d \right\}.$$

To formulate $\mathscr{S}(\kappa)$, we need to introduce a special norm of $A$, which is equivalent to the Frobenius norm up to a change of coordinates.

**Definition 6** (*Z*-weighted norm). Let $h_Z(A) = Z^{-1}AZ$, where $Z$ is the matrix defined in Theorem 3. Then the norm is defined as $\|h_Z(\cdot)\|_F = \|\cdot\|_F \circ h_Z$.

---

[1]For stochastic cases, $\hat{A}_d$ is consistently estimated by *Prediction Error Minimization* or *Maximum Likelihood* methods [1], and $\lim_{N \to \infty} \mathbb{E}(\hat{A}_d(N)) = \exp(hA)$. If only finite samples are available, we cannot obtain the exactly equivalent $\mathscr{E}(A, I, h, \mathscr{S}(\kappa))$ from data.

Since we assume that $\hat{A}_d$ is fixed, i.e., the data $X$ is not used in the optimization problems defined here, the matrix $Z$ is constant. One can observe that

$$\|h_Z(\hat{A})\|_F = \text{vec}(\hat{A})^T (Z^T \otimes Z^{-1})^T (Z^T \otimes Z^{-1}) \text{vec}(\hat{A})$$

is a proper $(Z^T \otimes Z^{-1})^T (Z^T \otimes Z^{-1})$-weighted vector norm in terms of $\text{vec}(\hat{A})$. Using $\|h_Z(\cdot)\|_F$ is on the one hand simplifying the analysis we conduct throughout this section, and on the other explicitly penalizes the imaginary part of the eigenvalues without "distorting" them through the transformation by $Z$.

Now we define $\mathscr{S}(\kappa)$ using the norm $\|h_Z(\cdot)\|_F$. The basic idea is that one should exclude such $A$'s whose imaginary parts of eigenvalues are too large, which implies their system response will show wild fluctuation. To make our assumption and the problem (11) practically meaningful, instead of $\mathbb{R}^{n \times n}$, we restrict $\mathscr{S}$ to be a norm bounded subset

$$\mathscr{S}(\kappa) = \left\{ \tilde{A} \in \mathbb{R}^{n \times n} : \|h_Z(\tilde{A})\|_F \leq \kappa \right\}. \qquad (13)$$

In the following we will show that the feasible set of (11) has only finite elements, which implies it can be solved at least by brutal force methods. Recall that the set $\mathcal{S}$ is countable according to Theorem 3.

Let $M := \text{diag}(m_1, m_2, \ldots, m_p)$, $\mathbf{j} := [j_1, j_2, \ldots, j_p]$ and $\beta := [\beta_1, \beta_2, \ldots, \beta_p]$, where $\log(\lambda_k) \triangleq \alpha_k + i\pi\beta_k$, $k = 1, \ldots, p$, and $j_k, \lambda_k$ are defined in Theorem 2. A function $\mathscr{I}$ is defined as

$$\mathscr{I}(\mathbf{j}, \delta) := \delta^T M \delta + (2\mathbf{j} + \beta)^T M \delta, \qquad (14)$$

where $\mathbf{j}, \delta \in \mathbb{Z}^p$. Moreover, it satisfies $\mathscr{I}(\mathbf{j}, \delta) = \mathscr{I}(0, \mathbf{j} + \delta) - \mathscr{I}(0, \mathbf{j})$, which follows by noticing

$$\mathscr{I}(\mathbf{j}, \delta) = (\delta + \mathbf{j} + \beta/2)^T M (\delta + \mathbf{j} + \beta/2)$$
$$- (\mathbf{j} + \beta/2)^T M (\mathbf{j} + \beta/2). \qquad (15)$$

Moreover, let $A_0$ denote a special matrix logarithm for which all $j_k$ ($k = 1, \ldots, p$) in (3) are equal to 0.

**Definition 7** (equivalence relations). An *equivalence* relation "$\sim$" is defined on $\mathcal{S}$ as a binary relation: for any $A_1, A_2 \in \mathcal{S}$, $\mathbf{j}^{(1)}$ and $\mathbf{j}^{(2)}$ are defined for $A_1, A_2$, respectively, we say $A_1 \sim A_2$ if $\mathscr{I}(\mathbf{j}^{(1)}, \mathbf{j}^{(2)} - \mathbf{j}^{(1)}) = 0$.

**Lemma 10.** *Let $\mathcal{S}$ be the set defined in (12) and parametrized by (4) in Theorem 3. For any $A_1, A_2 \in \mathcal{S}$, $\|h_Z(A_1)\|_F = \|h_Z(A_2)\|_F$ if and only if $A_1 \sim A_2$.*

*Proof.* Let $A_i := Z \text{diag}(L_1^{j_1^{(i)}}, \cdots, L_p^{j_p^{(i)}}) Z^{-1}$, where $i = 1, 2$, $L_k^{j_k^{(i)}} := \log(J_k(\lambda_k)) + 2j_k^{(i)} \pi i I_{m_k}$, and all other notations are given in (4). By using (16) (see next page) for $A_1, A_2$, we obtain

$$\|h_Z(A_1)\|_F = \|h_Z(A_2)\|_F \Leftrightarrow \|h_Z(A_1)\|_F^2 - \|h_Z(A_0)\|_F^2$$
$$= \|h_Z(A_2)\|_F^2 - \|h_Z(A_0)\|_F^2 \Leftrightarrow \mathscr{I}(\mathbf{j}^{(1)}, \mathbf{j}^{(2)} - \mathbf{j}^{(1)}) = 0,$$

which implies that $A_1 \sim A_2$ by definition. The first equality in (16) is due to the linear transformation $h_Z(\hat{A})$. $\square$

*Remark 3.* It is not necessary that $A_0$ is the principal matrix logarithm (consider the case when the principal logarithm does not exist), nor does it have to be the logarithm with the smallest (weighted) Frobenius norm.

**Lemma 11.** *Given any $\bar{A} \in \mathcal{S}$, there exist finite $A_i \in \mathcal{S}$ that satisfies $A_i \sim \bar{A}$.*

*Proof.* Let $\mathbf{j}$ denote $[j_1, \ldots, j_p]$ of $\bar{A}$ in (3), and $\mathbf{j}^{(i)}$ denotes $[j_1^{(i)}, \ldots, j_p^{(i)}]$ of $A_i \in \mathcal{S}$. $\delta \triangleq \mathbf{j}^{(i)} - \mathbf{j}$, therefore $\delta \in \mathbb{Z}^p$, where $\succeq$ denote the element-wise larger-or-equal relation. By Definition 7, it is equivalent to show that $\mathscr{I}(\mathbf{j}, \delta) = 0$ has finite solutions, given $\mathbf{j}$. We require $\delta$ to satisfy the following condition:

$$|\delta_i + j_i + \beta_i/2| \leq \sqrt{\frac{(\mathbf{j} + \beta/2)^T M (\mathbf{j} + \beta/2)}{m_i}} \qquad (17)$$

for all $i = 1, \ldots, p$. Otherwise, supposing that there exists $i \in \{1, \ldots, p\}$ such that $\delta_i$ does not satisfy (17), we will have

$$\mathscr{I}(\mathbf{j}, \delta) = m_i(\delta_i + j_i + \beta_i/2)^2 + \sum_{k \neq i} m_k(\delta_k + j_k \beta_k/2)^2$$
$$- \sum_k m_k(j_k + \beta_k/2)^2 > \sum_{k \neq i} m_k(\delta_k + j_k \beta_k/2)^2 \geq 0.$$

Let $\mathcal{S}' := \{A_i \in \mathcal{S} : j_k^{(i)} = \delta_k + j_k, \delta_k \text{ satisfies (17)}\}$. We have $\{A_i \in \mathcal{S} : A_i \sim \bar{A}\} \subseteq \mathcal{S}'$ and $\mathcal{S}'$ is a finite set. $\square$

**Lemma 12.** *There exists finite $A_i \in \mathcal{S}$ such that $\|h_Z(A_i)\|_F \leq \kappa$.*

*Proof.* Let $\kappa_0 \triangleq \|h_Z(A_0)\|_F$. Then we need to show there exists a finite number of $A_i \in \mathcal{S}$ such that $\|h_Z(A_i)\|_F^2 - \|h_Z(A_0)\|_F^2 \leq \kappa^2 - \kappa_0^2$, which is equivalent to show that there exists a finite number of solutions $\delta \in \mathbb{Z}$ to $\mathscr{I}(0, \delta) \leq (\kappa^2 - \kappa_0^2)/4\pi$. $\delta$ must satisfy the following condition:

$$|\delta_i + \beta_i/2| \leq \sqrt{\frac{(\beta/2)^T M (\beta/2) + (\kappa^2 - \kappa_0^2)}{m_i}} \qquad (18)$$

for all $i = 1, \ldots, p$. Otherwise, by supposing that there exists $i \in 1, \ldots, p$ such that $\delta_i$ does not satisfy (18) leads to

$$\mathscr{I}(0, \delta) = m_i(\delta_i + \beta_i/2)^2 +$$
$$\sum_{k \neq i} m_k(\delta_k + \beta_k/2)^2 - (\beta/2)^T M (\beta/2)$$
$$> \sum_{k \neq i} m_k(\delta_k + \beta_k/2)^2 + (\kappa^2 - \kappa_0^2) \geq \kappa^2 - \kappa_0^2.$$

Note that the set of all $\delta \in \mathbb{Z}$ that satisfies (18) is finite, which finalizes the proof. $\square$

*Remark 4.* We could have a more precise bound of $\delta_i$. Let $\mu(i) := \underset{\delta}{\text{minimize}} \sum_{k \neq i} m_k(\delta_k + \beta_k/2)^2$. Then $\delta_i$ ($i = 1, \ldots, p$) should satisfy

$$|\delta_i + \beta_i/2| \leq \sqrt{\frac{(\beta/2)^T M (\beta/2) + (\kappa^2 - \kappa_0^2) - \mu(i)}{m_i}}. \qquad (19)$$

$$\begin{aligned}
\|h_Z(A_i)\|_F^2 - \|h_Z(A_0)\|_F^2 &= \operatorname{tr}\left(\operatorname{diag}^*(L_1^{j_1^{(i)}},\cdots,L_p^{j_p^{(i)}})\operatorname{diag}(L_1^{j_1^{(i)}},\cdots,L_p^{j_p^{(i)}})\right) \\
&\quad - \operatorname{tr}\left(\operatorname{diag}^*(L_1^{(0)},\cdots,L_p^{(0)})\operatorname{diag}(L_1^{(0)},\cdots,L_p^{(0)})\right) \\
&= \sum_{k=1}^{p}\operatorname{tr}\left(L_k^{j_k^{(i)}*}L_k^{j_k^{(i)}} - L_k^{(0)*}L_k^{(0)}\right) \\
&= \sum_{k=1}^{p}\operatorname{tr}\left(2j_k^{(i)}\pi i\left(\log(J_k)^* - \log(J_k)\right) + 4\pi^2 j_k^{(i)2}I_{m_k}\right) \\
&= \sum_{k=1}^{p} 4\pi j_k^{(i)} m_k(\beta_k + j_k^{(i)}) = 4\pi\,\mathscr{I}(0,\mathbf{j}), \quad \mathbf{j} \triangleq [j_1^{(i)},\ldots,j_p^{(i)}].
\end{aligned} \qquad (16)$$

Moreover, we have the solution to $\mu(i), i = 1,\ldots,p$:

$$\mu(i) = \min\{\mathscr{I}(0,\delta_{/i}) : (\delta_{/i})_k = \lceil -\beta_k/2 \rceil \text{ or } \lfloor -\beta_k/2 \rfloor,\\ k \neq i;\ (\delta_{/i})_i = 0\}. \qquad (20)$$

**Proposition 13** (lower boundness of logarithms). *Let $\mathcal{S}$ be the set defined in* (12). *Given any $\bar{A} \in \mathcal{S}$, there exists $M(\bar{A}) > 0$, such that for any $A \in \{A \in \mathcal{S} : A \nsim \bar{A}\}$, it holds that*

$$\left|\|h_Z(A)\|_F - \|h_Z(\bar{A})\|_F\right| \geq M.$$

*Proof.* Let $\mathbf{j}$ denote $[j_1,\ldots,j_p]$ of $\bar{A}$ in (3), $N_{\text{eqiv}}$ be the number of $A$'s that satisfy $A \sim \bar{A}$. Note that $\left|\|h_Z(A)\|_F^2 - \|h_Z(\bar{A})\|_F^2\right| = \left|(\|h_Z(A)\|_F^2 - \|h_Z(A_0)\|_F^2) - (\|h_Z(\bar{A})\|_F^2 - \|h_Z(A_0)\|_F^2)\right| = |\mathscr{I}(\mathbf{j},\delta)|, \delta \in \mathbb{Z}$, which implies it is equivalent to show that $|\mathscr{I}(\mathbf{j},\delta)|, \delta \in \mathbb{Z}$ has a non-zero lower bound if not considering the $\delta$'s that result in $\mathscr{I}(\mathbf{j},\delta) = 0$. We will prove it by contradiction. Assume this is not true, i.e. $\forall \epsilon > 0$ there exists $\delta$ such that $0 < |\mathscr{I}(\mathbf{j},\delta)| < \epsilon$. It implies that, arbitrarily given $\epsilon > 0$, there exists an infinite number of $\delta$ such that $\mathscr{I}(\mathbf{j},\delta) < \epsilon$, which is impossible since $\mathscr{I}(0,\mathbf{j}+\delta) < \mathscr{I}(0,\mathbf{j}) + \epsilon$ (using the fact that $\mathscr{I}(\mathbf{j},\delta) = \mathscr{I}(0,\mathbf{j}+\delta) - \mathscr{I}(0,\mathbf{j})$) has a finite number of solutions provided by Lemma 12. $\square$

**Proposition 14.** *Let $\mathcal{S}$ be the set defined in* (12). *For any $\bar{A} \in \mathcal{S}$, there exist $\kappa_l, \kappa_u \in \mathbb{R}$ in $\mathcal{S}(\kappa_l,\kappa_u) = \{\tilde{A} \in \mathbb{R}^{n\times n} : \kappa_l \leq \|h_Z(\tilde{A})\|_F \leq \kappa_u\}$ such that* (11) *has a unique optimal point in the sense of the equivalence relation in Definition 7.*

*Proof.* It immediately follows by choosing

$$\kappa_l > \max\{0, \|h_Z(\bar{A})\|_F - M(\bar{A})\},$$
$$\kappa_u < \|h_Z(\bar{A})\|_F + M(\bar{A}),$$

where $M(\bar{A})$ is the lower bound on the gap between $\bar{A}$ and any $A \nsim \bar{A} \in \mathcal{S}$, defined in Theorem 13. $\square$

## V. Conclusions

This paper addresses identification of continuous-time dynamical systems with sparse topologies. The key assumption is that the sampling frequency is low. Under this assumption a realization/identification problem comes to surface, which has largely been overlooked in the community. First we propose the minimal sampling frequency that guarantees no system aliasing. Allowing system aliasing, one needs to search over a collection of matrix logarithms to find the sparsest one. We provide theoretical results for when a unique solution exists up to a finite equivalence class.

## Appendix
### Proof of Lemmas

*A. Proof of Lemma 5*

*Proof.* Suppose $E' \neq E$ and $E' + \bar{A} \in \mathscr{A}(n)$, $E' \in \mathscr{S}$. Using (8) and the fact that $e^{\bar{A}+E} \neq e^{\bar{A}+E'}$ when $\bar{A} + E' \in \mathscr{A}(n)$, we know that $f_{\bar{A}}(E) \neq f_{\bar{A}}(E')$. Now, since

$$\mathscr{L} \cap \ker(D) = \{0\}$$

it holds that $Df_{\bar{A}}(E) \neq Df_{\bar{A}}(E')$, otherwise $Df_{\bar{A}}(E) = Df_{\bar{A}}(E')$ and $\Pr_{\ker(D)}(f_{\bar{A}}(E)) \neq \Pr_{\ker(D)}(f_{\bar{A}}(E'))$, which means that

$$\mathscr{L} \supseteq \operatorname{span}\{f_{\bar{A}}(E') - f_{\bar{A}}(E)\} \subseteq \ker(D).$$

Now,

$$\|D\operatorname{vec}(\exp(\bar{A}+E)) - D\operatorname{vec}(\exp(\bar{A}+E'))\|_2$$
$$= \|D(f_{\bar{A}}(E) - f_{\bar{A}}(E'))\|_2 \neq 0.$$

$\square$

*B. Proof of Lemma 6*

*Proof.*
**Only if:** Suppose there is

$$E' \in (f_{\bar{A}}^{-1}(\ker(D)) - \{0\}) \cap \mathscr{S}.$$

By using the fact that $\mathscr{S}$ is a linear subspace, one can show that it holds that, $E^* + sE' \in \mathscr{E}_S(\bar{A},tE,D,\mathscr{S})$ for all $s$ if $E^* \in \mathscr{E}_S(\bar{A},tE,D,\mathscr{S})$. Now it is easy to show that (9) does not hold.

**If:** Let

$$\begin{aligned}
b(t) &= DK(\bar{A},0)\operatorname{vec}(tE), \\
q(t) &= D(\operatorname{vec}(e^A)) - D(\operatorname{vec}(e^{\bar{A}})) - b(t), \\
Q &= DK(\bar{A},0), \\
x &= \operatorname{vec}(\tilde{E}).
\end{aligned}$$

$\mathscr{S}$ is a linear subspace, let us assume its dimension is $k$. There is a full rank matrix $P \in \mathbb{R}^{n \times k}$ such that

$$\mathscr{S} = \{y : y = Pv, v \in \mathbb{R}^k\}.$$

The problem

$$\min_{\tilde{E} \in \mathscr{S}} \|D\mathrm{vec}(e^A) - D\mathrm{vec}(e^{\bar{A}}) - DK(\bar{A},0)\mathrm{vec}(\tilde{E})\|_2\},$$

can equivalently be written as

$$\begin{cases} \min & x^T Q^T Q x - 2(b(t) + q(t))^T Q x \\ \text{s.t.} & x = Pv. \end{cases}$$

The solution to this problem is

$$x = P(P^T Q^T Q P)^{-1} P^T Q^T (b(t) + q(t)).$$

Note that $P^T Q^T Q P = P^T K(\bar{A},0)^T D^T D K(\bar{A},0) P$ is invertible since $\mathrm{im}(K(\bar{A},0)P) \cap \ker(D) = 0$, $K(\bar{A},0)$ is invertible and $P$ has full rank. These facts combined with the fact that $P$ has no right-nullspace can be be used to show that $\mathscr{E}_S(\bar{A}, tE, D, \mathscr{S})$ contains only one element. It holds that $\mathrm{vec}(tE) = P(P^T Q^T Q P)^{-1} P^T Q^T (b(t))$. Also, $q(t)$ is $O(t^2)$; (9) holds. □